%% file: main.tex
\documentclass[conference]{IEEEtran}
\IEEEoverridecommandlockouts
\usepackage{cite}
\usepackage{amsmath,amssymb,amsfonts}
\usepackage{algorithm}
\usepackage{algorithmic}
\usepackage{graphicx}
\usepackage{textcomp}
\usepackage{xcolor}
\usepackage{url}
\usepackage{booktabs}
\usepackage{pifont}
\newcommand{\cmark}{\ding{51}}  
\newcommand{\xmark}{\ding{55}}  
\def\BibTeX{{\rm B\kern-.05em{\sc i\kern-.025em b}\kern-.08em
    T\kern-.1667em\lower.7ex\hbox{E}\kern-.125emX}}

\usepackage{listings}
\usepackage{xcolor}

\lstdefinestyle{bashstyle}{
  language=bash,
  basicstyle=\ttfamily\small,
  keywordstyle=\color{blue}\bfseries,
  commentstyle=\color{gray}\itshape,
  stringstyle=\color{orange},
  backgroundcolor=\color{white},
  showstringspaces=false,
  frame=single,
  breaklines=true,
  columns=fullflexible
}

\usepackage{array}
\usepackage{longtable}
\usepackage{multirow}

\begin{document}

\title{CodableLLM: Automating Decompiled and Source Code Mapping for LLM Dataset Generation}

\author{\IEEEauthorblockN{Dylan Manuel}
\IEEEauthorblockA{\textit{Department of Computer Science} \\
\textit{University of Texas at San Antonio}\\
San Antonio, TX, United States of America \\
dylan.manuel@utsa.edu}
\and
\IEEEauthorblockN{Paul Rad}
\IEEEauthorblockA{\textit{Department of Computer Science} \\
\textit{University of Texas at San Antonio}\\
San Antonio, TX, United States of America \\
peyman.najafirad@utsa.edu}
}

\maketitle

\begin{abstract}
The generation of large, high-quality datasets for code understanding and generation remains a significant challenge, particularly when aligning decompiled binaries with their original source code. To address this, we present \textbf{CodableLLM}, a Python framework designed to automate the creation and curation of datasets by mapping decompiled functions to their corresponding source functions. This process enhances the alignment between decompiled and source code representations, facilitating the development of large language models (LLMs) capable of understanding and generating code across multiple abstraction levels. \textbf{CodableLLM} supports multiple programming languages and integrates with existing decompilers and parsers to streamline dataset generation. This paper presents the design and implementation of \textbf{CodableLLM}, evaluates its performance in dataset creation, and compares it to existing tools in the field. The results demonstrate that \textbf{CodableLLM} offers a robust and efficient solution for generating datasets tailored for code-focused LLMS.
\end{abstract}


\section{Introduction}
\input{sections/1_introduction}

\section{Related Work and Motivation}
\input{sections/2_related_work_and_motivation}

\section{Design and Implementation}
\input{sections/3_design_and_implementation}

\section{Experiment and Evaluation}
\input{sections/4_experiment_and_evaluation}

\section{Discussion}
\input{sections/5_discussion}

\section{Conclusion}
\input{sections/6_conclusion}

\section{Licensing Attribution}

All source code used in this experiment was derived from the \texttt{libhv} project\footnote{\url{https://github.com/ithewei/libhv}}, licensed under the BSD 3-Clause License. The dataset and results are provided for academic purposes with full attribution to the original authors.

\bibliographystyle{IEEEtran}
\bibliography{bibliography}

\end{document}

%% file: sections/1_introduction.tex
The advancement of large language models (LLMs) has significantly impacted various domains, including natural language processing and code generation. Training these models requires substantial datasets that accurately represent the target domain. In the context of code understanding and generation, datasets that encompass both source code and its decompiled counterparts are invaluable. They enable models to learn the intricate relationships between high-level code and low-level representations, which is essential for tasks such as reverse engineering, vulnerability detection, and code transformation.

However, creating such datasets is a complex and time-consuming process. It involves decompiling binaries, extracting meaningful code segments, and aligning them with their original source code, all while navigating data duplication, quality control, and licensing challenges that have been shown to adversely impact dataset usability \cite{Allamanis2019Duplication,Improta2025Quality,Abt2014Synthetic}.

\begin{figure*}
    \centering
    \includegraphics[width=1\linewidth]{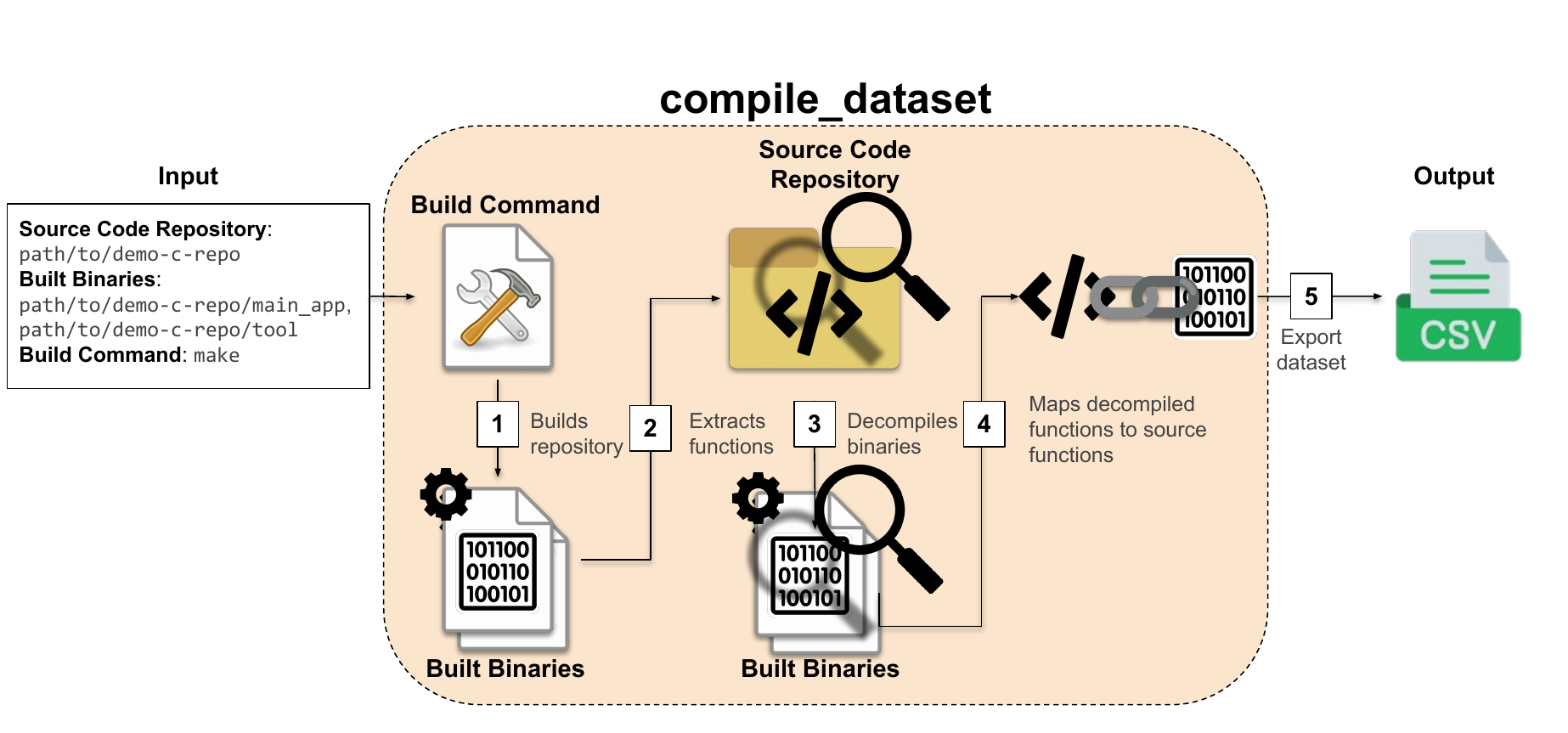}
    \caption{The main workflow of \textbf{CodableLLM}. In this example, an unbuilt local repository, \texttt{demo-c-repo}, is used as input to \textbf{CodableLLM}, specifying that the \texttt{make} command will be used to build the repository. The resulting binaries, \texttt{main\_app} and \texttt{tool}, are built from this process. \textbf{CodableLLM} first executes the build command to produce these binaries. It then extracts all source code functions from the repository's source files. Next, a decompiler is used to extract all functions from \texttt{main\_app} and \texttt{tool}. \textbf{CodableLLM} compares symbol names between the source and decompiled functions, maps them as pairs, and exports the resulting dataset as a CSV file.}
    \label{fig:codablellm_workflow}
\end{figure*}

To address this challenge, we introduce \textit{\textbf{CO}de \textbf{DA}taset \textbf{B}ui\textbf{L}d\textbf{E}r for \textbf{LLM}s} (\textbf{CodableLLM}), an extensible Python framework designed to automate the creation and curation of high-quality code datasets tailored for LLM training. \textbf{CodableLLM} streamlines the process of mapping decompiled binaries back to their source code, supporting multiple programming languages and integrating seamlessly with existing decompilers and parsers. By automating these tasks, \textbf{CodableLLM} significantly reduces the effort required to generate datasets, thereby facilitating the development of more capable code-focused LLMs. Figure~\ref{fig:codablellm_workflow} illustrates the high-level workflow of \textbf{CodableLLM}.

The main contributions of this work are summarized as follows:

\begin{itemize}
    \item We present \textbf{CodableLLM}, the first open-source framework specifically designed to automate the mapping between decompiled and source code for LLM dataset generation. \textbf{CodableLLM} features an extensible architecture that allows users to define custom parsing and function extraction logic for languages not supported by default.
    \item We demonstrate the framework's practical effectiveness across real-world repositories, highlighting its key strengths and limitations.
    \item We have open-sourced our code\footnote{\url{https://github.com/dmanuel64/codablellm/}}, published the framework as a package on the Python Package Index (PyPI)\footnote{\url{https://pypi.org/project/codablellm/}}, provided extensive documentation for users\footnote{\url{https://codablellm.readthedocs.io/en/latest/}}, and made the dataset used in our experiment publicly available\footnote{\url{https://github.com/dmanuel64/codablellm/paper/data/}}.
\end{itemize}

%% file: sections/2_related_work_and_motivation.tex
\subsection{Function Extraction and Parsing Tools}

Tools such as Tree-sitter \cite{tree_sitter} and SrcML \cite{srcML} provide language-agnostic parsing capabilities that facilitate the extraction of code functions and structural elements. These tools serve as foundational components for dataset creation pipelines. However, they are primarily designed for static code analysis rather than automated dataset curation or aligning source with decompiled representations.

\subsection{Existing Code Dataset Generation Efforts}

Large-scale datasets such as CodeSearchNet \cite{codesearchnet}, The Pile - Code \cite{the_pile}, datasets produced by the BigCode initiative \cite{bigcode}, and IBM's CodeNet \cite{Puri2021CodeNet} have significantly contributed to the availability of source code data for LLM training. These datasets encompass millions of functions across multiple programming languages and support a variety of downstream tasks, including code search, summarization, and generation. However, despite their scale and utility, these corpora are not without limitations. Recent studies have demonstrated that dataset bias and code duplication can lead to inflated model performance metrics and misleading conclusions about generalization capabilities \cite{Allamanis2019Duplication,Improta2025Quality}. In addition, acquiring large volumes of permissively licensed, diverse, high-quality code remains a persistent challenge, due in part to licensing restrictions and legal complexities \cite{Abt2014Synthetic,Raff2020Survey}. Furthermore, while these datasets focus on source code and its associated metadata, they do not address the complexities of aligning source functions with their corresponding decompiled representations — an alignment that is crucial for advancing research in binary analysis and reverse engineering.

\subsection{Binary Decompilation and Analysis Tools}

In the realm of binary analysis and reverse engineering, tools like Ghidra \cite{ghidra} and the Hex-Rays IDA Pro Decompiler \cite{ida_pro}, and machine learning-based disassembly frameworks such as XDA \cite{Pei2021XDA} are widely used. These tools translate binary executables into pseudo-C representations or other human-readable formats. However, even the most advanced tools often fall short of producing fully readable or semantically aligned source code representations \cite{Tan2024LLM4Decompile}. Furthermore, while highly effective for manual analysis and reverse engineering, they lack automation features for large-scale dataset extraction and do not provide pipelines for systematically aligning decompiled code with its original source.

\subsection{Motivation for CodableLLM}

Despite significant progress in code dataset development and binary analysis tooling, a critical gap remains in frameworks capable of automating the entire pipeline from binary decompilation to source-function alignment for large-scale dataset generation. Recent efforts, such as CAPYBARA \cite{AlKaswan2023Summarise}, have highlighted this deficiency in the context of binary function summarization, noting the absence of datasets that map decompiled functions to their original high-level code. Similarly, Assemblage \cite{Liu2024Assemblage} automates the generation of large binary datasets but stops short of supporting reverse mapping from binaries to source code. Other work, such as SourceFinder \cite{Rokon2020SourceFinder}, demonstrates attempts to correlate binaries with potential source repositories via large-scale code search; however, such approaches are domain-specific and not designed for generalized, multi-language dataset curation. To date, no open-source framework provides a configurable, extensible, and language-agnostic infrastructure for automating both decompiled and source code extraction, alignment, and dataset export at scale. \textbf{CodableLLM} addresses this gap by offering a unified and extensible framework that integrates decompilation, function extraction, heuristic-based mapping, and dataset creation, enabling research and model training that require linked binary and source code representations. Table~\ref{tab:related-tools} summarizes the key differences between \textbf{CodableLLM} and existing tools and datasets.

\begin{table*}[!t]
\centering
\begin{tabular}{|l||c|c|c|c|}
\hline
\textbf{Feature}                       & \textbf{Tree-Sitter} & \textbf{CodeSearchNet} & \textbf{Ghidra} & \textbf{CodableLLM} \\ \hline\hline
Source code parsing                    & \cmark               & \cmark                 & \xmark           & \cmark              \\ \hline
Multi-language support                 & \cmark               & \cmark                 & \xmark           & \cmark              \\ \hline
Decompiled function extraction         & \xmark               & \xmark                 & \cmark           & \cmark              \\ \hline
Automated source-to-decompiled mapping & \xmark               & \xmark                 & \xmark           & \cmark              \\ \hline
Dataset generation capabilities        & \xmark               & \cmark                 & \xmark           & \cmark              \\ \hline
Configurable build system integration  & \xmark               & \xmark                 & \xmark           & \cmark              \\ \hline
Open-source                            & \cmark               & \cmark                 & \cmark           & \cmark              \\ \hline
Extensible language support            & \xmark               & \xmark                 & \xmark           & \cmark              \\ \hline
\end{tabular}
\caption{Comparison of \textbf{CodableLLM} to existing tools and datasets.}
\label{tab:related-tools}
\end{table*}

%% file: sections/3_design_and_implementation.tex
\subsection{System Overview}

\begin{figure*}
    \small
    \centering
    \includegraphics[width=.9\linewidth]{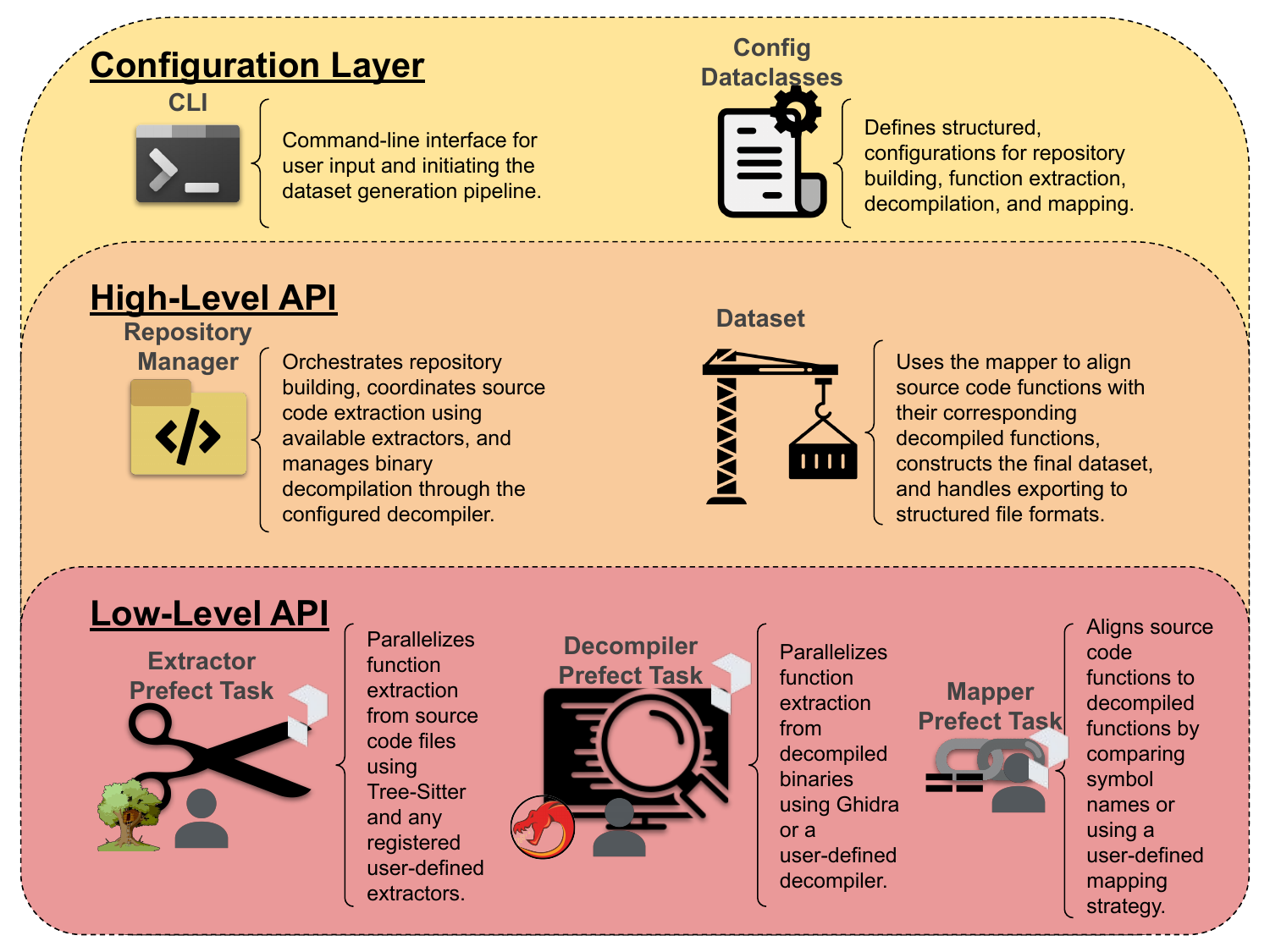}
    \caption{Layered architecture of \textbf{CodableLLM}, illustrating the configuration interface, high-level orchestration, and low-level Prefect-parallelized extraction and decompilation with support for user-defined extensions.}
    \label{fig:codablellm_arch}
\end{figure*}

\textbf{CodableLLM} is designed as a modular, extensible framework to facilitate the automated generation of datasets that align decompiled binaries with their corresponding source code functions. The architecture follows a multi-stage pipeline that accepts either local repositories or remote GitHub repositories as input, performs function extraction from both source code and decompiled binaries, aligns these functions by symbol names, and exports the results in structured formats suitable for large language model (LLM) training. The system prioritizes configurability, enabling users to specify build commands, repository paths, decompilers, and custom parsing logic through a unified interface. The layered architecture of \textbf{CodableLLM}, shown in Figure~\ref{fig:codablellm_arch}, highlights the modular design, parallel processing components, and extensibility through user-defined extractors and decompilers.

\subsection{Function Extraction Pipeline}

Function extraction from source code is performed using Tree-Sitter, an incremental parsing library that supports numerous programming languages. \textbf{CodableLLM} leverages Tree-Sitter's abstract syntax tree (AST) traversal capabilities to identify function-level constructs and extract metadata such as function names, parameters, and bodies. For languages not supported by Tree-Sitter, the framework allows users to implement custom extractors by adhering to a defined interface. This design decision ensures that the framework remains adaptable as new languages and parsing methodologies emerge.

\subsection{Binary Decompilation and Parsing}

The binary decompilation stage is integrated with Ghidra, a widely used open-source reverse engineering tool. Users can specify binaries to decompile, along with their respective paths and desired output formats. \textbf{CodableLLM} automates the execution of decompilation tasks and parses the resulting pseudo-C code to extract function-level information. The extracted decompiled functions are stored with metadata to facilitate later alignment with source code functions. Furthermore, the framework allows users to configure alternative decompilers or integrate proprietary tools if desired.

\subsection{Source-Decompiled Function Mapping}

Once both source and decompiled functions have been extracted, \textbf{CodableLLM} performs a symbol-based alignment process. This involves matching function names between the two domains and validating the correspondence based on configurable heuristics, such as filename consistency or namespace structures. In cases where symbol names are obfuscated or unavailable, the framework provides mechanisms for partial matching or exclusion. This alignment process is critical for generating datasets that can train models to reason across levels of abstraction.

\subsection{Dataset Export and Formats}

After mapping functions, \textbf{CodableLLM} exports the aligned pairs in CSV format by default. Each row contains fields such as repository name, file path, function name, source code body, decompiled code body, and optional metadata fields. Datasets within \textbf{CodableLLM} are constructed using the \texttt{pandas} \cite{pandas} DataFrame library, a widely adopted framework for efficient data manipulation and analysis. This design choice enables users to seamlessly export datasets to a variety of formats natively supported by DataFrames, including CSV, JSON, and other structured representations commonly used in machine learning workflows. Table~\ref{tab:dataset-fields} provides an overview of the fields included in a \textbf{CodableLLM} dataset, detailing the structure and metadata associated with each extracted function.

\subsection{Parallelism}

As of version 1.1.0, \textbf{CodableLLM} adopts \texttt{Prefect} \cite{prefect}, a Python-based workflow orchestration framework, to coordinate and parallelize each stage of the dataset generation pipeline. Prefect enables declarative task graphs, fault-tolerant execution, and seamless integration with Docker-based workflows. Each stage in the pipeline—such as source extraction, decompilation, mapping, and export—is encapsulated as a \texttt{Prefect} task, providing improved scheduling, observability, and resilience compared to traditional thread pool-based implementations. This design enhances extensibility while simplifying debugging and monitoring across large-scale dataset generation workflows.

\begin{table*}[!t]
\centering
\renewcommand{\arraystretch}{1.2}
\begin{tabular}{|l||p{12cm}|}
\hline
\textbf{Field} & \textbf{Description} \\
\hline\hline
\texttt{decompiled\_uid} & A unique identifier assigned to each decompiled function entry. \\
\hline
\texttt{assembly} & The disassembled assembly instructions corresponding to the decompiled function, extracted from the binary. \\
\hline
\texttt{architecture} & The target architecture (e.g., x86\_64, ARM) of the binary from which the function was decompiled. \\
\hline
\texttt{name} & The function name recovered or inferred from the decompiled binary symbol table or analysis. \\
\hline
\texttt{bin} & The file path to the binary from which the decompiled function was extracted. \\
\hline
\texttt{decompiled\_definition} & The full C-like function definition as produced by the decompiler. \\
\hline
\texttt{language} & The programming language of the original source code (e.g., C). \\
\hline
\texttt{source\_files} & A dictionary mapping each source function UID to the file path(s) where that function's source definition is located. \\
\hline
\texttt{source\_definitions} & A dictionary mapping each source function UID to the full source code definition of that function, enabling direct comparison with the decompiled definition. \\
\hline
\texttt{source\_file\_start\_bytes} & A dictionary mapping each source function UID to the starting byte offset of that function in its respective source file. \\
\hline
\texttt{source\_file\_end\_bytes} & A dictionary mapping each source function UID to the ending byte offset of that function in its respective source file. \\
\hline
\texttt{class\_names} & A dictionary mapping each source function UID to the fully qualified class or namespace name associated with that function, or \texttt{None} if not applicable (procedural code). \\
\hline
\end{tabular}
\caption{Description of dataset fields produced by \textbf{CodableLLM}. Fields related to source code mapping are represented as dictionaries keyed by source function unique identifiers (UIDs), allowing one-to-many relationships between source code definitions, locations, and associated metadata.}
\label{tab:dataset-fields}
\end{table*}

\subsection{Extensibility and User Customization}

A core design principle of \textbf{CodableLLM} is extensibility. Users can define custom build commands for language-specific compilation processes, integrate additional decompilers, and implement bespoke parsers for unsupported languages. The configuration system in \textbf{CodableLLM} is designed to be highly extensible, allowing users to define custom extractors and decompilers by subclassing existing components and registering them within the framework. This modular design enables researchers to adjust pipeline behavior and parameters without modifying the core codebase, ensuring that \textbf{CodableLLM} can accommodate diverse use cases and adapt to evolving research objectives.

\subsection{Algorithmic Overview}

The dataset generation process in \textbf{CodableLLM} follows a structured, multi-stage pipeline designed for scalability and extensibility. To accurately reflect the underlying architecture and facilitate reproducibility, Algorithm~\ref{alg:codablellm_pipeline} presents a detailed pseudocode representation of this pipeline. The process begins by cloning and building the target repository, followed by parallel extraction of source functions using available extractors and concurrent decompilation of specified binaries. The extracted functions are then mapped using heuristic matching logic implemented in the mapper module. Finally, the aligned function pairs are exported to structured dataset formats. This algorithm highlights the integration of user-defined extractors and decompilers, as well as the system's parallel processing capabilities.

\begin{algorithm}[!t]
\caption{CodableLLM Dataset Generation Pipeline}
\label{alg:codablellm_pipeline}
\begin{algorithmic}[1]
\REQUIRE $R$ (repository path), $B$ (build command), $b$ (list of target binaries)
\ENSURE Dataset file (CSV or JSON)

\STATE source\_function\_list $\leftarrow$ []
\STATE decompiled\_function\_list $\leftarrow$ []
\STATE mappings $\leftarrow$ []
\STATE
\STATE load\_user\_extractors()
\STATE load\_user\_decompilers()
\STATE clone\_repository($R$)
\STATE execute\_build($B$)
\STATE
\STATE \textbf{// Extract source functions in parallel}
\FORALL{source\_file in repository (in parallel)}
    \STATE functions $\leftarrow$ extractor.extract(source\_file)
    \FORALL{f in functions}
        \STATE source\_function\_list.add(f)
    \ENDFOR
\ENDFOR
\STATE
\STATE \textbf{// Decompile binaries and extract functions in parallel}
\FORALL{binary in $b$ (in parallel)}
    \STATE decompiled\_functions $\leftarrow$ decompiler.decompile(binary)
    \FORALL{f in decompiled\_functions}
        \STATE decompiled\_function\_list.add(f)
    \ENDFOR
\ENDFOR
\STATE
\STATE \textbf{// Map functions using mapper heuristics}
\FORALL{decompiled\_function in decompiled\_function\_list}
    \FORALL{source\_function in source\_function\_list}
        \IF{mapper.is\_potential\_match(source\_function, decompiled\_function)}
            \STATE mappings.add(source\_function, decompiled\_function)
        \ENDIF
    \ENDFOR
\ENDFOR
\STATE
\STATE export\_dataset(mappings, export\_format)
\end{algorithmic}
\end{algorithm}

%% file: sections/4_experiment_and_evaluation.tex
To evaluate the effectiveness and performance of \textbf{CodableLLM}, we generate datasets for a function name recovery experiment using the \texttt{libhv} \cite{libhv} project, a real-world C networking library licensed under the BSD 3-Clause license. Two datasets are produced: one from stripped binaries and another from non-stripped binaries. We then compare these results against a simulated single-threaded extraction and decompilation pipeline, representing a naive approach that does not leverage \textbf{CodableLLM}'s parallel processing capabilities.

\subsection{Environment Setup}

All experiments were conducted on a high-performance Linux server running Red Hat Enterprise Linux 8.10 (Ootpa). The server is equipped with dual-socket Intel(R) Xeon(R) Gold 6248 CPUs operating at 2.50 GHz, each with 20 cores and 2 threads per core, for a total of 80 logical processors. The system has 754 GB of RAM, with 721 GB available at runtime. All code extraction and decompilation tasks were executed using Python 3.12.8, and Ghidra version 11.0.1 was used as the decompiler backend. The \texttt{libhv} repository was built using a custom \texttt{build\_libhv.sh} script, shown in Listing~\ref{lst:build_libhv}, which automates the build process in accordance with the instructions provided in the \texttt{libhv} project's README.

\subsection{Prefect-Based Local Evaluation}

In addition to the experiments conducted on the high-performance server, we evaluated the Prefect-based version of \textbf{CodableLLM} (v1.1.0) using a local Windows machine. This experiment utilized a Docker Compose environment with a pre-configured PostgreSQL database and Prefect runtime. Due to Docker limitations on the main evaluation server, this experiment was conducted separately on a local system.

The local evaluation was conducted on a Windows 11 machine equipped with a 13th Gen Intel(R) Core(TM) i7-13700KF processor, featuring 16 physical cores and 24 logical threads. The system had 48 GB of RAM and was running Python 3.12.8. This setup was used in conjunction with Docker Compose and Prefect to orchestrate the dataset generation pipeline in a containerized environment.

\begin{lstlisting}[style=bashstyle, caption={Bash script used to build the libhv repository.}, label={lst:build_libhv}]
#!/bin/bash

cd "$(dirname "$0")"
SCRIPT_DIR_PATH=`pwd`

cd $SCRIPT_DIR_PATH/libhv
./configure
make
\end{lstlisting}

\subsection{Methodology}

We used \textbf{CodableLLM} to build the \texttt{libhv} repository, decompile binaries, and map decompiled functions to their source code counterparts. This experiment measures: 1.) extraction time, 2.) decompilation time, 3.) mapping time, 4.) dataset export time, 5.) overall pipeline time

\begin{figure*}
    \centering
    \includegraphics[width=.9\linewidth]{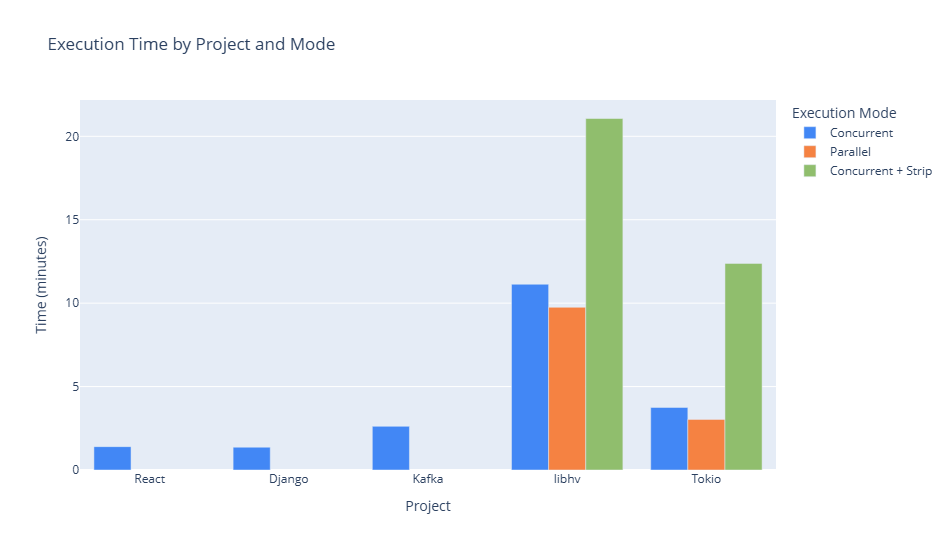}
    \caption{Execution time (in minutes) for function extraction and decompilation across multiple open-source repositories. Concurrent times were measured using multi-threaded execution. Only libhv and Tokio were additionally evaluated using a prior implementation of Prefect-based parallelism and a concurrent workflow with symbol stripping. The stripped mode reflects significantly longer processing time due to re-decompilation and reduced mapping efficiency.}
    \label{fig:pipeline_execution_time_comparison}
\end{figure*}

We also compare these results against a non-parallel single-extractor and single-decompiler simulation to demonstrate the performance benefits of \textbf{CodableLLM}'s thread pool implementation. Table~\ref{tab:codablellm-benchmark} and Figure~\ref{fig:pipeline_execution_time_comparison} highlight the performance of CodableLLM compared to other methods.

\begin{table*}[!t]
\scriptsize
\centering
\begin{tabular}{|l|l|r|r|r|r|}
\hline
\textbf{Project} & \textbf{Language(s)} & \textbf{Repo Size} & \textbf{Concurrent} & \textbf{Parallel} & \textbf{Concurrent + Strip} \\
\hline
React & JavaScript/TypeScript & 30.63 MB & 1m 24s & – & – \\
Django & Python & 42.98 MB & 1m 22s & – & – \\
Kafka & Java & 69.72 MB & 2m 37s & – & – \\
libhv & C/C++ & 3 MB & 11m 8s & 9m 45s & 21m 4s \\
Tokio & Rust & 5.17 MB & 6m 45s & 6m 2s & 12m 23s \\
\hline
\end{tabular}
\caption{Execution time by project using concurrent, parallel, and stripped binary workflows. Only \textbf{libhv} and \textbf{Tokio} include a Concurrent + Strip run, which applies symbol stripping prior to decompilation.}
\label{tab:codablellm-benchmark}
\end{table*}

%% file: sections/5_discussion.tex
\subsection{Strengths of \textbf{CodableLLM}}

\textbf{CodableLLM} introduces a fully automated pipeline for generating structured datasets that align source code functions with their decompiled counterparts. Unlike prior approaches that require manual intervention or ad hoc scripts, \textbf{CodableLLM} streamlines this process by integrating source extraction, decompilation, and function mapping into a single framework. The tool's extensible architecture enables researchers and practitioners to incorporate custom extractors and decompilers, making it adaptable to different programming languages and analysis needs.

One of the key strengths of \textbf{CodableLLM} is its use of parallelism to accelerate dataset generation. The evaluation results demonstrate that, compared to a naive single-threaded approach, \textbf{CodableLLM} reduces decompilation time from 514.52 seconds to 55.93 seconds, a nearly 10x improvement. Similarly, function mapping and dataset compilation are completed in under one second, highlighting the efficiency of its multi-threaded implementation. The Prefect-orchestrated pipeline executed the stripped \texttt{libhv} dataset configuration in 3 minutes and 19.01 seconds, compared to 3 minutes and 41.44 seconds for the previous thread pool-based implementation. Despite being run on significantly more constrained hardware, the Prefect pipeline exhibited reduced overhead and improved concurrency stability. These optimizations make \textbf{CodableLLM} practical for large-scale dataset generation, allowing researchers to construct datasets from real-world repositories without significant computational overhead.

Furthermore, the generated datasets are structured in standard formats such as CSV and JSON, making them easily integrable into downstream machine learning workflows. The ability to systematically generate and export aligned source-decompiled function pairs facilitates reproducibility and standardization, which are crucial for advancing research in decompilation, binary analysis, and AI-driven reverse engineering.

\subsection{Limitations}

Despite its advantages, \textbf{CodableLLM} has several limitations. First, its reliance on symbol-based mapping presents challenges when dealing with stripped binaries. In heavily optimized or obfuscated binaries, function names may be entirely removed, making direct alignment impossible. While \textbf{CodableLLM} currently supports basic mapping heuristics, more advanced techniques—such as function signature analysis or deep learning-based similarity models—are required to handle such cases effectively.

Second, the framework is currently dependent on Ghidra as its primary decompiler. While Ghidra offers a robust and open-source solution, it may produce inconsistent results in multi-threaded environments. Adding support for other tools such as Binary Ninja \cite{binary_ninja} and Radare2 \cite{radare2} provide alternative decompilation strategies that may yield better results in certain scenarios. Extending \textbf{CodableLLM} to support multiple decompilers would enhance its flexibility and applicability across diverse research and industry contexts. Additionally, as shown in Table~\ref{tab:codablellm-benchmark}, the single-threaded naive approach produced slightly more successful function mappings than the parallelized executions. This discrepancy is attributed to concurrency-related errors in Ghidra, highlighting the challenges of using Ghidra in multi-threaded environments and suggesting that improved concurrency handling or alternative decompilers may further benefit the framework.

Additionally, while the use of parallelism significantly improves performance, processing extremely large repositories may introduce memory overhead. Efficient memory management techniques, such as batch processing or streaming-based dataset generation, could mitigate this concern in future iterations of the framework.

\subsection{Implications}

The development of \textbf{CodableLLM} has several broader implications for research and industry. By providing an automated and scalable approach to dataset generation, \textbf{CodableLLM} lowers the barrier to entry for researchers working on binary-to-source function recovery, binary similarity detection, and AI-driven decompilation. This framework enables the rapid creation of structured datasets, which can be used to train and evaluate machine learning models for various reverse engineering tasks.

Moreover, \textbf{CodableLLM} facilitates reproducibility and standardization in decompilation-related research. Previously, many studies relied on manually curated datasets or proprietary tools, making it difficult to compare results across different methodologies. By providing an open-source, reproducible pipeline, \textbf{CodableLLM} helps establish consistent benchmarks for evaluating function recovery techniques.

Additionally, the tool has potential applications in cybersecurity, particularly in malware analysis and software vulnerability research. Automated function mapping could aid in identifying similarities between benign and malicious code, assisting in reverse engineering efforts for detecting obfuscated malware samples.

Finally, \textbf{CodableLLM} paves the way for benchmarking large language models (LLMs) on reverse engineering tasks. As LLMs continue to advance in code understanding and synthesis, evaluating their capabilities in function recovery from decompiled code remains an open research problem. The datasets generated by \textbf{CodableLLM} could serve as a foundation for such evaluations, providing high-quality training and testing data for AI-driven decompilation research.

\subsection{Future Work}

Future development of \textbf{CodableLLM} will focus on several key areas. First, 1.) advanced mapping heuristics will be introduced, incorporating structural similarity detection and potentially machine learning-based approaches to improve alignment accuracy for stripped or obfuscated binaries. Additionally, 2.) multi-decompiler support is planned, allowing integration with tools such as Binary Ninja and Radare2 to provide users with flexibility in choosing the decompiler most appropriate for their use case. Another priority is 3.) scalability and performance optimizations will be explored, focusing on memory-efficient batch processing techniques and dynamic resource management for handling extremely large repositories. Finally, 4.) visualization tools will be developed to assist researchers in interactively exploring mapped function pairs and validating dataset quality, making the dataset generation process more transparent and user-friendly.

%% file: sections/6_conclusion.tex
In this paper, we presented \textbf{CodableLLM}, the first open-source framework for automated source-to-decompiled function mapping and dataset generation.

We highlighted the system's extensible architecture, support for user-defined extractors and decompilers, and its parallel execution model, resulting in significant performance improvements over naive pipelines.

Our experiments on the libhv codebase demonstrate that \textbf{CodableLLM} is capable of generating high-quality datasets with substantial reductions in processing time.

By lowering the barrier for dataset generation in this domain, \textbf{CodableLLM} has broad implications for reverse engineering research, binary analysis, and the development of AI models for function recovery and decompilation.

Future work will continue to expand the framework's capabilities and extend its applicability across diverse architectures and toolchains, enabling standardized evaluation and reproducibility for the broader research community.